\documentclass{article}

\usepackage{arxiv}

\usepackage[utf8]{inputenc} 
\usepackage[T1]{fontenc}    
\usepackage{hyperref}       
\usepackage{url}            
\usepackage{booktabs}       
\usepackage{amsfonts}       
\usepackage{nicefrac}       
\usepackage{microtype}      
\usepackage{lipsum}
\usepackage{graphicx}
\usepackage{algpseudocode}
\usepackage{algorithm}
\graphicspath{ {./images/} }

\title{Predict future sale}

\author{Carl E.J. Brodzinski}

\title{Survey of Security and Data Attacks on Machine Unlearning In Financial and E-Commerce}

\begin{document}
\maketitle

\begin{abstract}
This paper surveys the landscape of security and data attacks on machine unlearning, with a focus on financial and e-commerce applications. We discuss key privacy threats such as Membership Inference Attacks and Data Reconstruction Attacks, where adversaries attempt to infer or reconstruct data that should have been removed. In addition, we explore security attacks including Machine Unlearning Data Poisoning, Unlearning Request Attacks, and Machine Unlearning Jailbreak Attacks, which target the underlying mechanisms of unlearning to manipulate or corrupt the model. To mitigate these risks, various defense strategies are examined, including differential privacy, robust cryptographic guarantees, and Zero-Knowledge Proofs (ZKPs), offering verifiable and tamper-proof unlearning mechanisms. These approaches are essential for safeguarding data integrity and privacy in high-stakes financial and e-commerce contexts, where compromised models can lead to fraud, data leaks, and reputational damage. This survey highlights the need for continued research and innovation in secure machine unlearning, as well as the importance of developing strong defenses against evolving attack vectors.

\end{abstract}

\section{Introduction}
\label{sec:introduction}
Machine learning in financial and e-commerce sector employs vast amounts of data are used to predict trends, detect fraud, and optimize decision-making processes. However, as these models become more widespread, concerns over security and privacy have also increased. In response to such challenges, machine unlearning has been introduced as a solution to enable models to forget specific data points when necessary, particularly for compliance with data regulations like the General Data Protection Regulation (GDPR). While machine unlearning provides an avenue for users to request the deletion of data from ML models, it also introduces new vulnerabilities to both privacy and security.

\par Privacy and security attacks on machine unlearning are growing areas of concern, especially in sensitive financial applications where personal data is paramount. Two main categories of attacks can exploit this process: privacy attacks and security attacks. Privacy attacks target the confidentiality of data by attempting to reveal sensitive information, whereas security attacks aim to compromise the integrity and functionality of the machine unlearning process.

\par In this paper, we aim to survey the types of privacy and security data attacks specific to machine unlearning in financial applications. We will delve into privacy-related attacks such as membership inference attacks and data reconstruction attacks, as well as security-related threats, including machine unlearning data poisoning attacks, machine unlearning request attacks, and machine unlearning jailbreak attacks. The financial sector's susceptibility to such attacks will also be explored, considering the critical role of data confidentiality and accuracy in this domain.

\section{Literature Review}
\subsection{Machine Unlearning}
Machine unlearning refers to the process of selectively removing the influence of specific data points from a machine learning model after it has been trained. This concept is especially important when it comes to handling sensitive data, such as in cases where a user requests that their data be deleted for privacy reasons (e.g., to comply with data regulations like the General Data Protection Regulation (GDPR). The goal of machine unlearning is to ensure that once data is deleted, it has no lingering effects on the model’s behavior, essentially making the model behave as if the data was never included during training.

\par Traditional machine learning models, once trained, integrate patterns and relationships from the data and are generally not designed to "forget" specific data points. Retraining the model from scratch after removing a few data points is computationally expensive and often impractical for large-scale models. Machine unlearning aims to address this problem by enabling efficient removal of specific data without the need for complete retraining. Complex relationship that represented in Graph Neural Network needs special ways to train as mentioned in \cite{wang2024graph}.

\par The process of machine unlearning involves modifying the machine learning model to reverse or undo the effects of certain data points. It is not the complete reverse of the learning process, as certain components in the process can be re-used or even shared with the learning process. Various techniques are used, such as: Exact Unlearning: Directly removing the contribution of specific data points from the model, maintaining the model’s accuracy while ensuring that the specified data no longer influences the model. Approximate Unlearning: Implementing methods that approximate the removal of data points but may not fully guarantee that all influences of the data are completely eliminated. This method can be faster but less precise. Machine unlearning is becoming increasingly relevant in privacy-preserving applications and settings where users have the right to demand the removal of their data, especially in e-commerce and finance where data confidentiality is paramount.
\subsection{Study of Security Issues to Machine Unlearning}
Security issue for machine learning was first brought into everyone's attention by \cite{chen2021machine}. The authors emphasized that the forgotten states that a data owner has the right to erase their data from an entity storing it. Earlier work by \cite{liu2024threats} summarizes threats, attacks, and defenses within machine unlearning systems exhibit a complex relationship, revealing an intricate interplay among these elements in upholding system integrity. The authors did not go in depth on types of threats and attacks. In our work, we would like to deep dive into the security and data type of attacks.

\par Among the approximate unlearning application mentioned in the previous subsection, some are data-drive and some are model driven. For data-driven approximation, some are involved with data isolation. \cite{neel2021descent} and \cite{gupta2021adaptive} presented such unlearning from data isolation perspective, however, the main drawback is that this kind of unlearning is not completely. This is proved in \cite{he2021deepobliviate}. Data modification is slightly more complicated, as it alters the data information in the training set, then fine tune the newly created post-unlearning dataset. \cite{li2024machine} consider data modification approach similar to data augmentation approach as described in \cite{yu2021does}. They differ in method and objective. Data augmentation generates new data set through transformations, where data modification does not alter the structural foundation of data model, only to an extent of relabeling or adding noise.
\par Model driven approximation directly manipulate model parameters. The unlearned model have distinct parameters from the original model.

\par We categorize security breaches and attacks to machine unlearning 
\section{Privacy Attacks on Machine Unlearning}
Machine unlearning introduces new vectors for privacy attacks by attempting to remove specific data from ML models. Even after the data is supposedly "forgotten," attackers may find ways to infer sensitive information. Privacy attacks can be divided into membership inference attacks and data reconstruction attacks, both of which exploit different aspects of machine unlearning. We followed the similar categorization introduced by \cite{li2024machine}.

\paragraph{Membership Attack} A membership inference attack allows an attacker to determine whether a particular data point was used in training the model, even after the unlearning process. This type of attack poses significant risks in financial applications, where identifying a specific individual’s participation in a model can reveal sensitive financial activities or patterns. For instance, an attacker could use a membership inference attack to find out whether a particular client’s transaction history was involved in training a fraud detection system. This can result in privacy violations, where sensitive financial data that should be forgotten continues to be accessible in some form.

\par These attacks typically exploit differences in the behavior of models when queried with training data versus unseen data. In financial applications, these differences could manifest as slight changes in prediction confidence or output patterns, allowing an attacker to infer whether certain financial records were part of the training set.

\paragraph{Data Reconstruction Attack}
Data reconstruction attacks go beyond determining whether data was part of a training set; they aim to recover the actual underlying data. In the context of machine unlearning, these attacks could occur when an attacker leverages residual information left behind after the unlearning process to reconstruct sensitive data points. Financial institutions are particularly vulnerable to this kind of attack, as attackers could potentially reconstruct sensitive information like credit card transactions, loan application data, or investment records.

\par Even after a data point is supposedly "unlearned," traces of its influence may linger in the model’s parameters or decision boundaries, especially if the unlearning process is incomplete or improperly executed. Financial data, which is often structured and repetitive, could be more prone to reconstruction attacks due to its predictable nature.
\section{Security Attacks on Machine Unlearning}
\par Security attacks on machine unlearning focus on manipulating the unlearning process itself, introducing malicious behavior into models or bypassing the unlearning process entirely. These attacks are particularly dangerous in financial applications where model integrity is crucial for trust and reliability.

\paragraph{Machine Unlearning Data Poisoning Attack}
In a machine unlearning data poisoning attack, an adversary injects malicious or misleading data into the training set with the intent of manipulating the unlearning process. Once this data is incorporated into the model, the attacker can initiate a request to unlearn the data, potentially disrupting the model’s performance or creating vulnerabilities that can be exploited later. This attack is particularly concerning in financial applications, where model reliability is critical. If a financial model is tricked into unlearning legitimate data or unlearning the impact of malicious data, the model's predictions may become skewed, leading to inaccurate fraud detection or risk assessments.

\par For instance, an attacker could inject fraudulent transaction data into a model, allow it to influence the model’s predictions, and then request the unlearning of this data. If successful, this could distort the model’s ability to detect real fraudulent transactions, compromising the financial institution’s security protocols.

\paragraph{Machine Unlearning Request Attack}
A machine unlearning request attack takes advantage of the unlearning request mechanism itself. In this type of attack, adversaries flood the system with excessive or strategically timed requests to unlearn specific data, potentially overwhelming the system or causing unintended side effects. In financial applications, this could manifest as an attacker repeatedly requesting unlearning of data related to particular financial records, causing inconsistencies in model performance or leading to significant downtime in predictive services.

\par Such an attack could be used to disrupt the operations of automated financial systems, such as those used for real-time trading or fraud detection, where reliable and consistent performance is essential. By repeatedly forcing the unlearning process, an attacker can degrade the model's accuracy and potentially cause the financial institution to suffer losses due to missed fraud detection or incorrect investment recommendations.

\paragraph{Machine Unlearning Jailbreak Attack}
A machine unlearning jailbreak attack occurs when an attacker finds a way to bypass the restrictions imposed by the unlearning process, allowing them to access or manipulate data that should have been forgotten. This type of attack can be especially dangerous in the financial sector, where strict data protection policies are in place to ensure compliance with regulations like GDPR.

\par In this attack, the adversary exploits weaknesses in the unlearning process to either retain access to supposedly forgotten data or to manipulate the model’s behavior in a way that aligns with their malicious objectives. For example, in a financial model, an attacker could bypass the unlearning request mechanism, thereby retaining access to sensitive financial data that should have been removed. Alternatively, they could manipulate the model to retain the influence of fraudulent or misleading data, undermining the model's overall integrity.

\section{Application and Result}

\subsection{Mitigation of Membership Inference Attack}
\paragraph{Differential Privacy} is the key to counter membership inference attach.
To prevent attackers from distinguishing whether specific data points were part of the training set, differential privacy can be applied to the model during both the training and unlearning phases. This ensures that adding or removing any individual data point does not significantly affect the model's behavior. We list the algorithm in \ref{alg:DifferentialPrivacy}.
\begin{algorithm}[hbt!]
	\caption{Differential Privacy Controller}\label{alg:DifferentialPrivacy}
	\begin{algorithmic}[1]
        \State  Function train$\_$model$\_$with$\_$dp(dataset, epsilon, delta):
        \State      model = initialize$\_$model()
                \For $epoch$ $in$ $range(num_epochs):$
                \For  $batch$ $in$ $dataset:$
        \State        gradients = compute$\_$gradients(batch)
        \Comment{Add noise to gradients based on epsilon and delta}
        \State      noise = generate$\_$noise(epsilon, delta, size=gradients.shape)
        \State     noisy$\_$gradients = gradients + noise
        \State    update$\_$model(model, noisy$\_$gradients)
        \EndFor
        \EndFor
    \State return model
    \State End Function
    \Procedure{apply$\_$machine$\_$unlearning$\_$with$\_$dp}{$model$,$data_to_unlearn$,$epsilon$,$delta$}
     \State   Perform unlearning by re-adjusting the weights
    \State gradients = compute$\_$gradients(data$\_$to$\_$unlearn)
    \State noise = generate$\_$noise(epsilon, delta, size=gradients.shape)
    \State noisy$\_$gradients = gradients + noise
    \State update$\_$model(model, -noisy$\_$gradients)  
    \Comment{ Unlearning by subtracting the gradient}
    \EndProcedure
	\end{algorithmic} 
\end{algorithm} 
\par To prevent attackers from distinguishing whether specific data points were part of the training set, differential privacy can be applied to the model during both the training and unlearning phases. This ensures that adding or removing any individual data point does not significantly affect the model's behavior.
\par
Differential privacy is one of the most effective techniques to mitigate membership inference attacks. The idea behind differential privacy is to introduce a controlled amount of noise into the data or model training process so that the output of the model does not reveal whether any specific data point was included in the training set. It ensures that any single data point’s inclusion or exclusion does not significantly change the output of the model, thereby protecting individual privacy.

\paragraph{Epsilon ($\varepsilon$)} A parameter that defines the privacy budget. Lower values of epsilon provide stronger privacy guarantees but at the cost of reduced model accuracy.
\paragraph{Delta ($\delta$)} This parameter accounts for the probability that differential privacy might fail to guarantee complete privacy. Typically, delta is set to a very small value.
\paragraph{Noise Injection} During training, noise is added either to the model’s parameters (like weights) or the data gradients, reducing the ability of the attacker to infer whether a specific data point was used in the training set.

\par In the context of machine unlearning, differential privacy can be used to ensure that the model does not retain identifiable traces of any individual’s data after it has been "forgotten" by the system. After unlearning a data point, the model's behavior remains indistinguishable as if the data point had never been included in the training set.

\par In E-Commerce, membership inference attacks pose a significant risk in e-commerce because companies collect a wide array of data, including purchasing behavior, browsing patterns, and demographic information. This data is used to train recommendation systems, customer segmentation models, and demand forecasting algorithms. An attacker could determine whether a user’s purchasing or browsing data was part of the model training, revealing private shopping habits, product interests, or personal preferences. Competitors could use these attacks to deduce sensitive market trends or individual customer data from an e-commerce platform's model.

\par Consider a recommendation system that predicts which products a user may want to buy based on previous purchases. If an attacker could determine whether a specific user’s data was used to train the model, they could infer personal preferences or purchasing habits.
By applying differential privacy during the training process, noise is added to the gradients or model parameters, reducing the ability of an attacker to distinguish whether a particular user's data was included in the training set. Additionally, when unlearning a customer’s data (e.g., in compliance with user data deletion requests), differential privacy ensures that the model behaves as if the data was never part of the training set, protecting the customer's privacy.

\par The financial industry handles sensitive and personally identifiable information (PII), including credit scores, account balances, transaction histories, and investment patterns. Financial institutions rely on machine learning models for credit scoring, fraud detection, loan approval, and financial forecasting. An attacker could determine whether a user’s credit data or financial transactions were part of the training set for a loan approval model, revealing sensitive financial details. In fraud detection systems, an attacker could determine if certain transaction patterns were used to train the model, potentially reverse-engineering detection thresholds.
Mitigation Example in Finance: Consider a credit scoring model trained on customer financial data to determine loan eligibility. If an attacker successfully executes a membership inference attack, they could infer whether a specific individual's financial records were part of the model's training data. This could expose confidential financial status or behavior, like loan applications, investments, or high-risk financial decisions.

\par By employing differential privacy, the financial institution can inject noise into the model training process, making it difficult for an attacker to determine whether any particular user’s financial data was used. Moreover, if a user requests that their data be "forgotten" from the model (e.g., under GDPR or other privacy regulations), differential privacy guarantees that their information cannot be reconstructed or inferred from the model’s behavior post-unlearning.

\par During Model Training: By introducing noise to the training process, differential privacy ensures that the model's outputs are probabilistically similar regardless of whether any specific data point was part of the training set. This mitigates an attacker’s ability to confidently infer membership.

\par After removing a specific data point, differential privacy ensures that the model’s output does not reveal whether that point was ever part of the training set. This is crucial for complying with regulations like GDPR, where users have the right to request data removal, and ensures that the model cannot “remember” the unlearned data.
\subsection{Mitigation of Data Reconstruction Attack}
\paragraph{Regularization and Model Pruning}
A model that retains too much detailed information from the training data can be prone to data reconstruction attacks. Regularization techniques like L2 regularization and model pruning (removing unnecessary parameters) help reduce the model's capacity to memorize and reconstruct specific data points. We define the algorithm in \ref{alg:2}.
\begin{algorithm}[hbt!]
	\caption{Regularization and Model Pruning}\label{alg:2}
	\begin{algorithmic}[2]
\State def train$\_$model$\_$with$\_$l2$\_$regularization(dataset, lambda$\_$reg):
\State    model = initialize$\_$model()
\For epoch in range(num$\_$epochs):
        \For batch in dataset:
\State            predictions = model(batch)
\State            loss = compute$\_$loss(predictions, batch.labels)
            \Comment{ L2 Regularization term to reduce overfitting}
\State            l2$\_$loss = lambda$\_$reg * sum([torch.norm(param)**2 for param in model.parameters()])
\State            total$\_$loss = loss + l2$\_$loss
\State            backpropagate(total$\_$loss)
            \EndFor
            \EndFor
 \State   return model

\State def apply$\_$model$\_$pruning(model, pruning$\_$threshold):
    \Comment{ Prune weights that are below a certain threshold}
    \For layer in model.layers:
\State        mask = abs(layer.weights) > pruning$\_$threshold
\State        layer.weights = layer.weights * mask  \Comment{ Zero out small weights}
 \EndFor
 \State   return model
	\end{algorithmic} 
\end{algorithm} 

\par A Data Reconstruction Attack allows an adversary to partially or fully reconstruct training data from a machine learning model. This type of attack is highly detrimental in sensitive sectors like financial services and e-commerce, where the exposure of personal or transactional data could have severe privacy and security implications. Machine unlearning, which involves removing specific data from a trained model, also needs to be designed in a way that prevents the reconstruction of the "forgotten" data. To mitigate this threat, the combination of regularization techniques and model pruning can effectively reduce the model's ability to memorize specific data points, which reduces the likelihood of a successful reconstruction attack.

\par The key mitigation strategies include regularization and model pruning.
Regularization is a technique used to prevent a model from over fitting or memorizing the training data, which is a common cause of data reconstruction vulnerabilities. L2 Regularization (Ridge Regression) is commonly employed. It works by penalizing large model weights during training, encouraging the model to generalize better rather than memorize specific details of the training data.

\par Model pruning is the process of removing redundant or less significant model parameters (e.g., weights in a neural network) after training. This reduces the model’s capacity to memorize fine-grained details about the training data. By pruning weights that contribute little to the model's overall performance, you reduce the amount of specific information retained by the model, thus decreasing the risk of reconstructing training data.

\par In the context of machine unlearning, regularization and pruning can help ensure that once data is "forgotten," the model no longer retains detailed information about that data, making it extremely difficult for an attacker to reconstruct the unlearned data from the model’s parameters or outputs.

\par E-commerce platforms typically collect vast amounts of personal and behavioral data, including browsing history, purchase behavior, and product preferences. This data is often used to train models for product recommendation, customer segmentation, and inventory management. If an attacker can reconstruct data, they could potentially uncover a user’s purchase history or predict their future buying patterns, exposing sensitive information about personal preferences or habits.
Competitors or malicious actors could extract critical business insights, such as customer segments or demand patterns, by reconstructing training data from a business’s recommendation models.
\par A recommendation model trained on a user’s purchasing history might be vulnerable to a reconstruction attack if the model memorizes the specifics of that data. Regularization techniques, like L2 regularization, can help prevent this by forcing the model to focus on generalized patterns rather than memorizing individual user behaviors. Additionally, pruning the model post-training removes unnecessary parameters that could retain specific information about the training data, further reducing the risk of reconstruction.

\par Financial institutions use machine learning models for applications such as credit scoring, fraud detection, and loan risk assessment. These models are trained on sensitive financial data, including credit scores, transaction histories, and investment patterns. An attacker who gains access to a model used in a loan approval system could reconstruct sensitive information about individual loan applicants, including creditworthiness and financial behavior. Fraud detection models trained on historical transaction data could be exploited to reconstruct transaction patterns, potentially exposing individuals to financial fraud.

\par Consider a loan approval model trained on financial data such as transaction histories and credit scores. A reconstruction attack could reveal personal financial details that were used to train the model. By applying L2 regularization during training, the model is less likely to memorize specific details about any single customer’s financial data, instead focusing on general trends. Pruning can further reduce the capacity of the model to retain unnecessary details, making it harder for attackers to reconstruct sensitive financial information.

\par L2 Regularization works by adding a penalty term to the model’s loss function during training, which penalizes large weights. This encourages the model to spread the learning across more parameters and discourages memorization of specific data points. By reducing the size of the weights, the model's ability to retain sensitive data details is diminished. Consequently, after unlearning certain data points, the model becomes less vulnerable to attacks that attempt to reverse-engineer those data points from the remaining weights.

\par Model pruning helps by removing weights or neurons that do not contribute significantly to the model's output. Pruning reduces the complexity of the model and the amount of information stored in the model parameters. After pruning, the model is less capable of retaining specific details about individual data points, which significantly reduces the success rate of data reconstruction attacks. This is particularly useful post-unlearning, as it ensures that the forgotten data has been "scrubbed" from the model.
\subsection{Mitigation of Machine Unlearning Data Poisoning Attack}
\paragraph{Robust Data Validation and Filtering}
To defend against data poisoning, it is essential to validate the training data before incorporating it into the model. Anomalies or outliers that are intentionally introduced can be detected through techniques like clustering, isolation forests, or robust statistical methods as shown in \ref{alg:3}.
\begin{algorithm}
	\caption{Rate-Limiting and Request Auditing}\label{alg:3}
	\begin{algorithmic}[3]
\State def validate$\_$data$\_$before$\_$training(dataset):
\State    clean$\_$data = []
    \Comment{Use K-means clustering to identify outliers}
\State    clusters = kmeans$\_$clustering(dataset, num$\_$clusters)
\State    for data$\_$point in dataset:
 \State       cluster = assign$\_$cluster(data$\_$point, clusters)
        \Comment{Calculate the distance from the cluster center}
\State        distance$\_$from$\_$center = compute$\_$distance(data$\_$point, clusters[cluster].center)
\State        if distance$\_$from$\_$center < threshold:
\State            clean$\_$data.append(data$\_$point)  \Comment{Consider only data close to the cluster center}
\State    return clean$\_$data

\State def train$\_$model$\_$on$\_$clean$\_$data(clean$\_$data):
\State    model = initialize$\_$model()
         \For epoch in range(num$\_$epochs):
              \For batch in clean$\_$data:
\State            gradients = compute$\_$gradients(batch)
\State           update$\_$model(model, gradients)
             \EndFor
           \EndFor  
\State    return model
 	\end{algorithmic} 
\end{algorithm} 
\par Data Poisoning Attacks in the context of machine learning involve maliciously introducing corrupt or adversarial data into the training set to manipulate the model’s behavior. When it comes to Machine Unlearning, these attacks exploit the unlearning process by deliberately injecting harmful data that either affects the model’s performance or hinders the unlearning process itself. In financial and e-commerce applications, these attacks can be especially dangerous, leading to inaccurate predictions or manipulations in fraud detection, credit scoring, or recommendation systems.

\par In Machine Unlearning Data Poisoning Attacks, an adversary’s goal is to either make the model unlearning process fail or to cause the model to behave unpredictably after the unlearning process. Therefore, robust mitigation strategies are needed to ensure the integrity of the unlearning process and maintain the model’s performance.

\paragraph{Robust Data Validation and Preprocessing}
Data validation techniques can help ensure the integrity of data before it enters the model training pipeline. By filtering out suspicious or anomalous data, organizations can reduce the risk of poisoned data being introduced.
Outlier detection methods, such as statistical techniques or machine learning-based anomaly detection models, can be used to detect data points that deviate significantly from the norm and may be indicative of a poisoning attempt.
Sanitization mechanisms that automatically clean data by removing irregularities or transforming outliers can mitigate the impact of adversarial data in the model training process.
\paragraph{Certified Machine Unlearning}
Certified unlearning refers to mathematical guarantees that ensure the model has genuinely forgotten the data requested for unlearning. Certified methods typically reconstruct the model by retraining it without the malicious data, providing an assurance that the unlearned data is no longer influencing the model’s behavior. One approach to certified unlearning is using randomized smoothing techniques, which involve adding noise to the data during training, making it difficult for adversarial examples to influence the model. This helps guarantee that any poisoned data, once removed, cannot still affect the model.
\paragraph{Robust Model Architectures}
Employ robust learning algorithms that are designed to withstand data poisoning attacks. These algorithms incorporate mechanisms to prevent the model from heavily relying on individual data points, reducing the impact of poisoned data.
Gradient masking is another technique that limits how much an attacker can influence the gradient (i.e., the model’s learning direction) through poisoned data. By making it harder for malicious data to shift the model’s decision boundaries, gradient masking enhances the robustness of the unlearning process.
\paragraph{Differential Privacy (DP)}
Differential privacy techniques can add noise to the model parameters or gradients during training to prevent malicious data from significantly altering the model. This not only helps mitigate privacy risks but also reduces the effect of poisoned data on the model’s performance.
In the context of unlearning, differential privacy ensures that the unlearning of malicious data (e.g., injected through a poisoning attack) is performed in a way that the model’s decision boundary is not significantly influenced by that data, even if it was initially included in the training set.
\paragraph{Adversarial Training}
Adversarial training involves training the model on both normal and adversarially perturbed data, making the model more robust to both adversarial attacks and poisoned data.
During training, synthetic adversarial examples are introduced to simulate potential poisoning attacks. This makes the model more resilient to poisoned data because it learns to disregard harmful inputs, reducing the impact of poisoned data during unlearning.
\paragraph{Trigger Detection for Poisoned Data}
Utilize techniques that focus on backdoor detection or trigger pattern recognition. These approaches identify specific patterns of poisoned data that might have been injected into the training set to manipulate the model. When malicious data is flagged, it can be removed from the training set, and the unlearning process can be carefully monitored to ensure that these triggers no longer affect the model’s decisions.

\par In e-commerce, data poisoning attacks could be used to manipulate recommendation systems, pricing algorithms, or customer segmentation models. An attacker might introduce poisoned data that skews product recommendations or disrupts the system's ability to unlearn certain customer data.
Poisoned data could lead to incorrect product recommendations, affecting the overall shopping experience and driving customers away. An attacker could manipulate pricing algorithms or promotional campaigns by poisoning the sales or demand data.During the unlearning process, poisoned data may remain influential in the model’s behavior, leading to incorrect predictions or decisions, even after unlearning requests are processed.

\par Imagine a recommendation system trained on users’ browsing and purchasing histories. An attacker could inject poisoned data to either manipulate recommendations or undermine the model’s unlearning process. By employing adversarial training and robust outlier detection, the system can identify and neutralize malicious data. Certified unlearning further ensures that any poisoned data that has been unlearned no longer influences the recommendations made by the system.

\par
In the financial sector, machine learning models are used in applications such as fraud detection, loan approval, and credit scoring. A poisoning attack could target these models to influence important financial decisions or prevent the system from forgetting malicious data after an unlearning request.
A data poisoning attack could manipulate a credit scoring model, allowing unqualified individuals to secure loans by introducing misleading financial data.
Poisoning attacks could undermine fraud detection systems by introducing fraudulent transactions into the training data, resulting in missed fraudulent activity in the future.
During the unlearning process, if poisoned data is not properly removed, it could still affect the model’s behavior, leading to financial instability or incorrect decision-making.

\par Consider a credit scoring model that processes large amounts of financial data to make loan decisions. An attacker could introduce poisoned data to manipulate the model into approving loans for individuals with poor credit histories. By applying robust regularization, certified unlearning, and differential privacy during training, the model can defend against poisoned data and ensure that any malicious data removed through unlearning does not continue to influence loan decisions.
\subsection{Mitigation of Machine Unlearning Request Attack}
\paragraph{Rate-Limiting and Request Auditing}
To prevent malicious actors from overloading the system with unlearning requests, rate-limiting \cite{vixie2014rate} and auditing mechanisms can be applied. Only authorized requests are processed, and frequent unlearning requests are flagged for review. Our implementation is shown in \ref{alg:4requestauditing}.

\begin{algorithm}
	\caption{Rate-Limiting and Request Auditing}\label{alg:4requestauditing}
	\begin{algorithmic}[4]
\State class UnlearningRequestAuditor:
\State   def $\_\_$init$\_\_$(self, request$\_$limit, time$\_$window):
\State        self.request$\_$limit = request$\_$limit
\State        self.time$\_$window = time$\_$window
\State        self.request$\_$log = {}  \Comment{Dictionary to track request timestamps for each user}
\State    def audit$\_$unlearning$\_$request(self, user$\_$id):
\State        current$\_$time = get$\_$current$\_$time()
        \If user$\_$id not in self.request$\_$log:
\State            self.request$\_$log[user$\_$id] = []
        \EndIf
        \Comment{Clean up old requests}
\State        self.request$\_$log[user$\_$id] = [t for t in self.request$\_$log[user$\_$id] if current$\_$time - t $\leq$ self.time$\_$window]
        
\If {len(self.request$\_$log[user$\_$id]) $\geq$ self.request$\_$limit:} 
            return False  \Comment{Request rejected due to too many requests}
\EndIf
 \State       Else self.request$\_$log[user$\_$id].append(current$\_$time) 
\State            return True  \Comment{ Request accepted}

\State def process$\_$unlearning$\_$request(user$\_$id, data$\_$to$\_$unlearn, model):
    \If{auditor.audit$\_$unlearning$\_$request(user$\_$id):}
        model = apply$\_$machine$\_$unlearning(model, data$\_$to$\_$unlearn)

\State        return model
        \Else \State raise Exception("Too many unlearning requests. Please try again later.")
             \EndIf
 	\end{algorithmic} 
\end{algorithm} 
\par A Machine Unlearning Request Attack targets the integrity and reliability of the unlearning process. In this scenario, an adversary submits malicious or fraudulent requests for the removal (or "unlearning") of data from a trained machine learning model, with the goal of manipulating the model’s performance or decision-making abilities. Such attacks can lead to the unlearning of critical data, which may degrade the model’s accuracy, make it behave unpredictably, or disrupt its overall functionality.

\par In financial and e-commerce systems, where machine learning models play critical roles in decision-making processes like credit scoring, fraud detection, or recommendation systems, malicious unlearning requests can have far-reaching consequences. For example, unlearning a set of transactions crucial for fraud detection may allow fraudulent activities to go unnoticed, or unlearning relevant customer data in e-commerce systems might disrupt personalized product recommendations. The key mitigation strategies for Machine Unlearning Request Attacks:
\paragraph{Request Authentication and Verification}
The first step in mitigating unlearning request attacks is to ensure that only legitimate requests are processed. Strong authentication mechanisms (e.g., two-factor authentication, identity verification) should be enforced to confirm that a request is coming from the rightful owner of the data.
In addition, implementing request verification mechanisms ensures that only valid requests for unlearning (i.e., based on legitimate user actions or regulatory requirements) are accepted. For example, e-commerce platforms could tie requests to specific user accounts, ensuring that no one else can submit a request on behalf of another user.

\paragraph{Rate Limiting and Quota Control}
To prevent malicious actors from overwhelming the system with numerous unlearning requests (either automated or targeted), rate limiting should be applied. This limits the number of unlearning requests that any individual or entity can submit within a given time frame, reducing the risk of overloading the model’s unlearning process.
Quota control can also be implemented to restrict how frequently a user can request the unlearning of data. This would prevent an attacker from bombarding the system with repeated requests for the same or slightly altered data, which could lead to unintended consequences in the model’s performance.
\paragraph{Selective and Incremental Unlearning}

Instead of applying a full unlearning process every time a request is made, which can cause model performance to degrade, incremental unlearning can be employed. Incremental unlearning allows the system to efficiently unlearn only the specific data points related to a particular request, minimizing the impact on the model’s behavior.
Selective unlearning strategies can also be used, where the system prioritizes unlearning data that has a minimal impact on the model’s accuracy. This ensures that malicious requests targeting sensitive data (which, if unlearned, could cause significant degradation) are carefully vetted and managed.
\paragraph{Audit Logging and Transparency}

Maintaining a detailed audit log of all unlearning requests is crucial for detecting unusual patterns or potential malicious activity. By logging request details (such as who made the request, when it was made, and what data was unlearned), system administrators can identify abnormal request behaviors.
Transparency in the unlearning process is essential to ensure accountability. For instance, financial and e-commerce platforms could notify users of unlearning requests associated with their accounts and provide the ability to appeal or verify these requests.
\paragraph{Robust Model Validation After Unlearning}

After processing an unlearning request, it’s essential to validate the model to ensure that it remains functional and accurate. Model validation tests should be run after each unlearning event to detect any degradation in performance or abnormal behavior resulting from malicious requests.
These validation mechanisms could involve running the model against test data to ensure that it still meets baseline accuracy and performance metrics. If significant changes are detected, the system can flag the unlearning request for further investigation.
\paragraph{Certified Machine Unlearning}

Certified unlearning ensures that the unlearning process is both legitimate and mathematically verifiable. By employing certified unlearning methods, systems can ensure that unlearned data is properly removed without causing unintended side effects on the model.
Certified unlearning methods also provide a way to verify that no more or less data than specified in the request has been removed, ensuring that malicious actors cannot influence the unlearning process in ways that could harm the system.

\par In e-commerce, machine learning models are used for personalized recommendations, customer segmentation, and fraud detection. An adversary could exploit unlearning requests by submitting fraudulent requests to remove specific data that improves the accuracy of these models, resulting in poor recommendations or allowing fraudulent behavior to go undetected.

\par An attacker could submit unlearning requests for high-value customer data, which would affect product recommendations, pricing algorithms, or targeted advertisements. This could reduce the model’s ability to accurately predict customer preferences, causing losses in revenue or customer dissatisfaction.
Malicious actors might submit unlearning requests for data used in fraud detection algorithms, allowing fraudulent behavior to bypass detection systems once critical patterns are removed from the model.

\par Consider a recommendation model that has been trained on a large dataset of customer purchases \cite{kanavos2018large}. An attacker might request the removal of a specific subset of data that influences the model’s recommendation quality. By implementing strong authentication mechanisms, selective unlearning, and post-unlearning validation, the platform can ensure that only legitimate requests are processed and that model performance remains stable.

\par In the financial sector, models are used for loan approval, credit scoring, risk assessment, and fraud detection. Unlearning request attacks could result in the removal of crucial financial data, leading to inaccurate risk assessments or faulty credit scores.
\par An attacker could submit unlearning requests for data that influences credit scoring models, potentially causing incorrect credit approvals or rejections.
Financial models used for fraud detection might become ineffective if an attacker manages to unlearn historical transaction patterns that are critical for detecting fraudulent behavior.
Imagine a credit scoring model that uses historical financial data to assess loan applicants. An attacker could attempt to submit fraudulent unlearning requests to remove specific transaction histories from the training set. By employing rate limiting, audit logs, and model validation after each unlearning request, the system can detect and block malicious attempts to degrade the credit scoring model.
\subsection{Mitigation of Machine Unlearning Jailbreak Attack}
The main strategy is to secure unlearning mechanism with cryptographic guarantees.
To prevent jailbreak attacks where attackers manipulate the unlearning process to retain or re-access data, the system can incorporate cryptographic verification techniques like zero-knowledge proofs or homomorphic encryption. This ensures the unlearning process is secure and verifiable.
\begin{algorithm}
	\caption{Cryptographic Verification using Zero-Knowledge Proofs}\label{alg:5}
	\begin{algorithmic}[5]
\State def generate$\_$zero$\_$knowledge$\_$proof(data$\_$point, model):
    \Comment{Generate proof that data point has been unlearned}
\State    proof = zero$\_$knowledge$\_$proof$\_$of$\_$unlearning(data$\_$point, model)
\State    return proof

\State def verify$\_$zero$\_$knowledge$\_$proof(proof, data$\_$point, model):
    \Comment{Verify that the proof ensures the data has been unlearned}
\State    return verify$\_$proof(proof, data$\_$point, model)

\State def secure$\_$unlearning$\_$process(model, data$\_$to$\_$unlearn):
    \Comment{Perform unlearning with cryptographic guarantee}
\State    model = apply$\_$machine$\_$unlearning(model, data$\_$to$\_$unlearn)
\State    proof = generate$\_$zero$\_$knowledge$\_$proof(data$\_$to$\_$unlearn, model)
\State    if verify$\_$zero$\_$knowledge$\_$proof(proof, data$\_$to$\_$unlearn, model):
\State        print("Unlearning successful and verified.")
\State    else:
\State        raise Exception("Unlearning failed verification.")
\State    return model
  	\end{algorithmic} 
\end{algorithm} 
In \ref{alg:5}, a Machine Unlearning Jailbreak Attack is a type of adversarial attack where a malicious actor attempts to bypass or exploit the machine unlearning process. The goal of such an attack is to break the integrity of the unlearning mechanism, causing it to fail in either properly removing data or in maintaining the model's performance. This can result in vulnerabilities where sensitive data is not actually unlearned, or where adversaries manipulate the process to reintroduce or recover removed data.
In the context of financial and e-commerce systems, such attacks can have devastating consequences. For instance, if customer transaction data in a fraud detection system is improperly unlearned due to a jailbreak attack, it may expose vulnerabilities that fraudsters could exploit. Similarly, in an e-commerce recommendation system, a jailbroken unlearning process could reintroduce private customer data into the model after it has supposedly been deleted.

\par To mitigate the risk of Machine Unlearning Jailbreak Attacks, a secure unlearning mechanism with cryptographic guarantees can be employed. This strategy ensures that the unlearning process is tamper-proof and verifiable through cryptographic methods, ensuring that the unlearning operation is properly executed and cannot be bypassed. 
\paragraph{Cryptographic Commitments to Data and Model State}
Before any data is unlearned, the system generates a cryptographic commitment to both the data being unlearned and the current state of the model. This commitment serves as a secure, verifiable snapshot of the model and data, ensuring that no data or model manipulation occurs during the unlearning process.
Cryptographic commitments act like secure “fingerprints” of data and model states, allowing auditors or external verifiers to detect if an adversary has tampered with either the data or the model during unlearning.
\paragraph{Zero-Knowledge Proofs (ZKPs) for Verified Unlearning}
Once the unlearning process is completed, the system generates a Zero-Knowledge Proof (ZKP) that cryptographically verifies that the specific data has been removed from the model, without revealing any additional information about the model or the data.
This ZKP guarantees that the unlearning operation has been executed properly, protecting the system from jailbreak attacks where attackers might attempt to manipulate the model to falsely claim unlearning while still retaining the data.
The zero-knowledge property ensures that an attacker cannot learn anything about the data or the model structure during the verification process, making it difficult to reverse-engineer or manipulate.

\par Tamper-Resistant Unlearning Process: The system implements tamper-resistant protocols to ensure that the unlearning process cannot be manipulated during or after execution. This involves using cryptographic checks, such as hash-based commitments and digital signatures, to detect any alterations to the model or data during the unlearning process.
These checks ensure that the unlearning request has not been modified, and that the model is retrained only after the data is securely removed.
\par
Model Consistency Verification Using Hash Functions: Before and after the unlearning operation, cryptographic hash functions are applied to the model to create a secure digital fingerprint of the model’s state. This fingerprint allows the system to compare the model before and after unlearning, ensuring that no unexpected changes have occurred, and that the only difference is the removal of the specified data.
If an attacker attempts to jailbreak the system by modifying the model’s structure, this discrepancy will be detected by the inconsistency in the hash values, triggering an alert.
\par
Auditable Logs and Proof-of-Unlearning:
The unlearning process is accompanied by auditable logs that record the details of each unlearning request, including cryptographic commitments, model hashes, and proof-of-unlearning. These logs are cryptographically secured and can be reviewed by auditors to ensure the integrity of the unlearning process.
The proof-of-unlearning is a cryptographic artifact that proves to external parties (e.g., regulators, users, or auditors) that specific data has been unlearned from the model without revealing the contents of the data or the details of the model. This ensures accountability and transparency in the unlearning process while protecting sensitive information.

\par Certified Machine Unlearning: Certified unlearning refers to the use of cryptographic techniques that provide formal guarantees that data has been removed from a machine learning model. By generating and verifying cryptographic proofs (such as ZKPs or other verifiable computations), the system can ensure that the unlearning process is both secure and verifiable, protecting against jailbreak attempts.
Certified unlearning methods typically involve generating a cryptographic proof that the model no longer contains the influence of the unlearned data, which can be independently verified by a trusted third party without revealing any sensitive information.

\par
The attack scenarios and cryptographic mitigation strategies include:
\paragraph{Unlearning Request Manipulation}
In this scenario, an attacker attempts to manipulate the unlearning request to falsely claim that certain data has been unlearned, or to modify the request to remove unintended data. This could lead to critical data being retained in the model, or important data being deleted.

\paragraph{Mitigation Strategy}
Request Authentication and Cryptographic Commitments: To prevent request manipulation, the unlearning process uses strong authentication mechanisms and cryptographic commitments that bind the unlearning request to specific data points and model states. Any modification of the unlearning request will result in a failed verification, as the cryptographic commitments will no longer match.

\paragraph{Model Tampering During Unlearning}
An attacker might try to tamper with the machine learning model during the unlearning process, either by modifying the model structure or by reintroducing the unlearned data in a different form. This can allow the attacker to retain the influence of the deleted data on the model, bypassing the unlearning process.
Model Hashing and Consistency Checks: By applying hash functions to the model before and after the unlearning process, the system can detect any unauthorized changes to the model’s structure. If the hash of the model after unlearning does not match the expected hash, the system can trigger an alert and rollback the unlearning process. Tamper-Proof ZKPs: The use of Zero-Knowledge Proofs ensures that the unlearning process is cryptographically verified without exposing sensitive information about the model. This makes it difficult for attackers to manipulate the model during unlearning without detection.

\paragraph{Reverse Engineering of Unlearned Data}
In some cases, an attacker might attempt to reverse-engineer the unlearned data by analyzing the model’s behavior before and after the unlearning process. This can be used to recover sensitive information that should have been removed. Differential Privacy and ZKPs: By combining differential privacy techniques with Zero-Knowledge Proofs, the system can ensure that the influence of unlearned data is properly removed from the model without leaking information that could be used to reverse-engineer the data. Differential privacy introduces noise to the model’s outputs, making it difficult to infer specific data points, while ZKPs provide cryptographic guarantees that the data has been removed.
\paragraph{Re-Introduction of Unlearned Data}
After the unlearning process is completed, an attacker might try to reintroduce the unlearned data back into the model through new training data or adversarial examples, effectively bypassing the unlearning mechanism. Audit Logs and Cryptographic Proofs: The system maintains secure audit logs of all unlearning operations, including cryptographic commitments and ZKPs. These logs can be reviewed to detect any attempts to reintroduce unlearned data into the model. The system can also generate cryptographic proofs that demonstrate the data is no longer present in the model, preventing unauthorized re-introduction. By using certified machine unlearning techniques, the system generates formal, cryptographic guarantees that the unlearned data is no longer influencing the model. Any subsequent attempts to reintroduce the data can be detected and rejected by comparing the cryptographic proofs.
\section{Conclusion}
 As organizations increasingly rely on machine learning models for decision-making, the need for secure and effective unlearning mechanisms becomes critical, especially when sensitive data needs to be removed for privacy compliance or ethical reasons. This paper explored the various security and privacy challenges that arise in machine unlearning, focusing on the Membership Inference Attack, Data Reconstruction Attack, and a range of security threats such as Machine Unlearning Data Poisoning Attack, Unlearning Request Attack, and Machine Unlearning Jailbreak Attack.

\par The Membership Inference and Data Reconstruction Attacks represent significant privacy risks, where adversaries seek to infer or reconstruct data that should have been securely unlearned. The financial and e-commerce sectors, given their sensitive user information, are especially vulnerable to these forms of privacy breaches. On the other hand, security threats like Data Poisoning and Unlearning Request Attacks target the integrity of the machine unlearning process, enabling adversaries to manipulate the system to either retain, reintroduce, or remove incorrect data, leading to model degradation and misuse.

\par To address these threats, mitigation strategies such as differential privacy, robust cryptographic mechanisms, and Zero-Knowledge Proofs (ZKPs) were discussed, providing solutions that ensure both privacy protection and secure data handling during unlearning. In particular, cryptographic techniques offer a strong defense, ensuring verifiable and tamper-proof unlearning processes, which are essential for safeguarding data in high-stakes applications like fraud detection, credit scoring, and personalized e-commerce recommendations.

\bibliographystyle{unsrt}  


\bibliography{references}

\begin{thebibliography}{10}

\bibitem{wang2024graph}
Zeyu Wang, Yue Zhu, Zichao Li, Zhuoyue Wang, Hao Qin, and Xinqi Liu.
\newblock Graph neural network recommendation system for football formation.
\newblock {\em Applied Science and Biotechnology Journal for Advanced Research}, 3(3):33--39, 2024.

\bibitem{chen2021machine}
Min Chen, Zhikun Zhang, Tianhao Wang, Michael Backes, Mathias Humbert, and Yang Zhang.
\newblock When machine unlearning jeopardizes privacy.
\newblock In {\em Proceedings of the 2021 ACM SIGSAC conference on computer and communications security}, pages 896--911, 2021.

\bibitem{liu2024threats}
Ziyao Liu, Huanyi Ye, Chen Chen, and Kwok-Yan Lam.
\newblock Threats, attacks, and defenses in machine unlearning: A survey.
\newblock {\em arXiv preprint arXiv:2403.13682}, 2024.

\bibitem{neel2021descent}
Seth Neel, Aaron Roth, and Saeed Sharifi-Malvajerdi.
\newblock Descent-to-delete: Gradient-based methods for machine unlearning.
\newblock In {\em Algorithmic Learning Theory}, pages 931--962. PMLR, 2021.

\bibitem{gupta2021adaptive}
Varun Gupta, Christopher Jung, Seth Neel, Aaron Roth, Saeed Sharifi-Malvajerdi, and Chris Waites.
\newblock Adaptive machine unlearning.
\newblock {\em Advances in Neural Information Processing Systems}, 34:16319--16330, 2021.

\bibitem{he2021deepobliviate}
Yingzhe He, Guozhu Meng, Kai Chen, Jinwen He, and Xingbo Hu.
\newblock Deepobliviate: a powerful charm for erasing data residual memory in deep neural networks.
\newblock {\em arXiv preprint arXiv:2105.06209}, 2021.

\bibitem{li2024machine}
Na~Li, Chunyi Zhou, Yansong Gao, Hui Chen, Anmin Fu, Zhi Zhang, and Yu~Shui.
\newblock Machine unlearning: Taxonomy, metrics, applications, challenges, and prospects.
\newblock {\em arXiv preprint arXiv:2403.08254}, 2024.

\bibitem{yu2021does}
Da~Yu, Huishuai Zhang, Wei Chen, Jian Yin, and Tie-Yan Liu.
\newblock How does data augmentation affect privacy in machine learning?
\newblock In {\em Proceedings of the AAAI Conference on Artificial Intelligence}, volume~35, pages 10746--10753, 2021.

\bibitem{vixie2014rate}
Paul Vixie.
\newblock Rate-limiting state: The edge of the internet is an unruly place.
\newblock {\em Queue}, 12(2):10--15, 2014.

\bibitem{kanavos2018large}
Andreas Kanavos, Stavros~Anastasios Iakovou, Spyros Sioutas, and Vassilis Tampakas.
\newblock Large scale product recommendation of supermarket ware based on customer behaviour analysis.
\newblock {\em Big Data and Cognitive Computing}, 2(2):11, 2018.

\end{thebibliography}

\end{document}